\documentclass[12pt,a4paper]{article}
\usepackage[latin1]{inputenc}
\usepackage{amsmath}
\usepackage{amsfonts}
\usepackage{amssymb}
\usepackage{graphicx}
\usepackage{hyperref}
\usepackage{authblk}

\title{Transverse Spin with Coupled Plasmons}

\author[1]{Samyobrata Mukherjee}
\author[2]{A V Gopal}
\author[1,*]{S Dutta Gupta}

\affil[1]{School of Physics, University of Hyderabad, Hyderabad, India}
\affil[2]{DPCMS, TIFR, Mumbai, India}
\affil[*]{Corresponding author: sdghyderabad@gmail.com}
\begin{document}
	\maketitle
	
	\begin{abstract}
		We study theoretically the transverse spin associated with the eigenmodes of a thin metal film embedded in a dielectric. We show that the transverse spin has a direct dependence on the nature and strength of the coupling leading to two distinct branches for the long- and short- range modes. We show that the short-range mode exhibits larger extraordinary spin because of its more 'structured' nature due to higher decay in propagation. In contrast to some of the earlier studies, calculations are performed retaining the full lossy character of the metal. In the limit of vanishing losses we present analytical results for the extraordinary spin for both the coupled modes. The results can have direct implications for enhancing the elusive transverse spin exploiting the coupled plasmon structures. 
	\end{abstract}
	

	
\section{Introduction}

It is now well understood that light fields carry both orbital and spin angular momentum (SAM) in addition to the conventional linear momentum. Together they play a vital role in light-matter interaction.  Since the pioneering work of Allen \cite{allen1992} spin-orbit interactions with light fields have become a popular area of research \cite{allen2000,aiello2009,bliokhreview1,bliokhreview2,aielloreview}. Recent investigations have revealed that structured light can exhibit an unusual transverse (perpendicular to propagation vector) SAM, which is independent of the helicity \cite{bliokh2014, prl_expt}. These extraordinary features can be probed in the evanescent fields of, say, surface plasmons at dielectric-metal interfaces \cite{bliokh2012}, and can result in the optical spin-momentum locking \cite{aiello2009,bliokhreview2,bliokh2014,bliokh2015,zayats2014,jacob2016}. The extraordinary SAM, first introduced by Belinfante \cite{belinfante1940} has been labelled as a virtual (elusive) entity, since it has negligible effect on dipolar particles. Recent studies on optical currents embodied by the Poynting vector have shown that the current consists of contributions of both the orbital and the spin momentum \cite{berry2009,berry2015}. Moreover, the transverse SAM was shown to be proportional to $Im\left(\mathbf{E}^*\times\mathbf{E}\right)$. For structured fields, finite longitudinal component of the fields (as in the case of surface plasmons) leading to a polarization ellipse in the plane of incidence is the source of the transverse spin momentum. Of late, there have been several investigations into Belinfante's ``elusive" transverse spin \cite{belinfante1940} in the context of structured light fields. It is now understood that this transverse spin can be both helicity dependent \cite{nat_expt} as well as helicity independent \cite{bliokh2012}. Multiple schemes have been proposed to probe the manifestation of this transverse spin \cite{bliokh2014, bliokh2012, bliokh2015, bekshaev2015, aiello2015, alizadeh2016} in order to gain better insight into the phenomenon.  There have also been reports of both direct and indirect experimental detection of this transverse spin \cite{prl_expt,nat_expt}. The direct measurement \cite{nat_expt} measured the torque on a nano-cantilever and the value measured was of the order of femtonewton. The miniscule nature of the quantity being measured makes it imperative to search for a means of enhancing this quantity. A recent study has shown that it is possible to exploit coherent perfect absorption \cite{wan,shourya,nireekshan,sdg2012} mediated mode enhancement in a gap plasmon structure in order to enhance transverse spin \cite{mukherjee2016}. In another study the localized plasmons and the Mie resonances were shown to be effective tools to enhance the transverse spin \cite{nirmalya2016}. However, there are no studies as yet on the most elementary form of coupled plasmons as in a thin metallic film.

In this paper we focus our attention on the eigenmodes of a thin metallic film and calculate the transverse spin for both the long and short range modes. Note that the coupled modes of a thin film has been the subject of extensive research for past several decades and the field enhancement associated with the long range mode has been identified as an effective tool to enhance the nonlinear optical effects \cite{sarid1982,sdg1986,sdg1987,sdg1990}. We show that for sufficient coupling the transverse spin splits into short range and long range branches. One may expect larger transverse spin due to larger local field enhancement for the long range mode. Contrary to this expectation, the transverse spin exhibited by the short range mode is shown to be larger than that of the long range one. This is ascribed to the more structured (with larger decay) nature for the short range mode. Our study thus reveals the importance of having structured fields for enhancing the transverse spin. 
\par
The structure of the paper is as follows. In section 2 we present a brief description of the system and formulate the problem. Following \cite{sdgbook} we present a simple matrix method to calculate the dispersion characteristics and the eigenmodes of the structure and use them to calculate the transverse spin. Section 3 contains the results of the numerical calculations retaining the losses  in the metal film. In contrast, in most of the existing studies intrinsic losses in the metal are ignored \cite{bliokh2014,bliokh2012}. We show distinct branches for the transverse spin corresponding to the bifurcation of the dispersion results. In section 4, we present closed-form expressions for the transverse spin for both the branches of the coupled modes ignoring the losses in the metal in order to gain theoretical insight. Major findings of this paper are summarized in the conclusion.

\section{Formulation of the Problem}

\begin{figure}[t]
	\centering
	\includegraphics[width=0.6\linewidth]{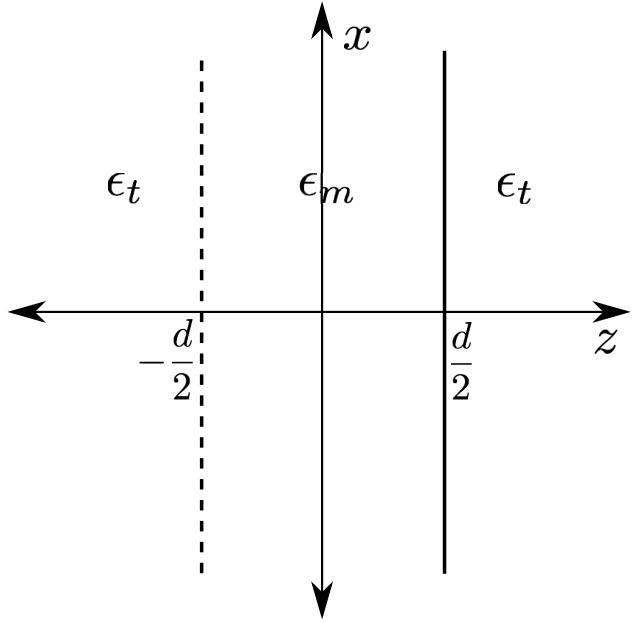}
	\caption{Schematic view of the geometry under consideration. We exploit the symmetry and consider only the right half of the structure.}
	\label{Structure_Geometry}
\end{figure}

Consider the simplest possible structure supporting coupled surface plasmons, as in the early work of Sarid \cite{sarid1982,sdgbook,raetherbook} (see figure \ref{Structure_Geometry}). It comprises of a thin metal film with dielectric constant $\epsilon_m$ of thickness $d$ placed in a dielectric with dielectric constant $\epsilon_t$. The structure is symmetric about the $x-y$ plane and we exploit this symmetry to simplify our calculations considering only the positive $z$ half-space. Since surface plasmons can  be excited only by TM polarised waves, we write the magnetic field inside the metal as
\begin{equation}\label{eq1}
\sqrt{\mu_{0}}H_{my} = \sqrt{\mu_{0}} A_{0}\left(e^{ik_{mz}z}\pm e^{-ik_{mz}z}\right) e^{ik_{x}x}\ .
\end{equation}
where $A_0$ is the amplitude of the magnetic field in the metal, $k_{mz} = i\sqrt{k_{x}^{2} - k_{0}^{2}\epsilon_{m}}= i\bar{k}_{mz}$.  One can arrive at similar equations for the fields in ambient dielectric where the  $z$-component of the wave vector $k_{tz} = i\sqrt{k_{x}^{2} - k_{0}^{2}\epsilon_{t}}= i\bar{k}_{tz}$. Note that we have put the $z$- component of the wave vectors in a form to highlight the evanescent character of the waves in both metal and the dielectric. Recall that we are looking at an eigenproblem where there are no external excitations and we are posing the question when the eigenmodes can exist and can propagate along $x$ axis with a given $z$ spatial profile localized near the film.
The upper and lower signs in eq. (\ref{eq1}) are for the symmetric and the antisymmetric modes, respectively. Here the symmetry is decided by the even-odd character of the magnetic spatial field distribution $H_{y}$ about the $z=0$ plane. 
Later we will refer to these modes as long-range and short range ones, judging by their propagation losses (imaginary part of $k_x$).
Using Maxwell's equations, we can calculate the field $E_{mx}$ as follows
\begin{equation}\label{eq2}
\sqrt{\epsilon_{0}}E_{mx} = p_{mz}\sqrt{\mu_{0}}A_{0}\left(e^{ik_{mz}z}\mp e^{-ik_{mz}z}\right) e^{ik_{x}x}\ ,
\end{equation}

Making use of the transfer matrices \cite{sdgbook}, we can relate the tangential components at $z=0$ and $z=\frac{d}{2}$ to obtain
\begin{equation}\label{eq3}
\left(\begin{array}{cc}
1 & \pm 1 \\
p_{mz} & \mp p_{mz} \end{array}\right)
\left(\begin{array}{c}
\sqrt{\mu_{0}}A_{0} \\
\sqrt{\mu_{0}}A_{0} \end{array}\right) = 
M_{T}\left(\begin{array}{c}
1\\
p_{tz} \end{array}\right)\sqrt{\mu_{0}}A_{t}\ ,
\end{equation}
where $A_{t}$ is the amplitude of $H_{y}$ in the dielectric, $p_{mz} = \frac{k_{mz}}{k_{0}\epsilon_{m}}$, $p_{tz} = \frac{k_{tz}}{k_{0}\epsilon_{t}}$. For our geometry (see figure \ref{Structure_Geometry}), the properties of the metal film are encoded in the transfer matrix 
\begin{equation}\label{eq4}
M_{T} = \left( \begin{array}{cc} 
\cos\left(\frac{k_{mz}d}{2}\right) & -\frac{i}{p_{mz}}\sin\left(\frac{k_{mz}d}{2}\right) \\
-ip_{mz}\sin\left(\frac{k_{mz}d}{2}\right) & \cos\left(\frac{k_{mz}d}{2}\right)	
\end{array}\right).
\end{equation}
As mentioned earlier, we get the symmetric and antisymmetric cases, by picking the upper and lower signs, respectively in eq. (\ref{eq3}). The transfer matrix equation (see eq. (\ref{eq3})) gives us the dispersion relation as well as the amplitude relation for both the coupled modes.
%
%
Using the upper sign in eq. \ref{eq3} we get the well known dispersion relation \cite{sdgbook,raetherbook} for the \textbf{symmetric} case
\begin{equation}\label{eq5}
\epsilon_{m}k_{tz} + \epsilon_{t}k_{mz} \tanh\left(\frac{\bar{k}_{mz}d}{2}\right) = 0\ .
\end{equation}		
We can now use this dispersion relation to derive the amplitude of the transmitted magnetic field in the dielectric
\begin{equation}\label{eq6}
A_{t}^{s} = 2 A_0 \cosh\left(\frac{\bar{k}_{mz}d}{2}\right)\ .
\end{equation}
Using the lower sign, we get the dispersion relation and transmitted amplitude for the \textbf{antisymmetric} case \cite{sdgbook,raetherbook}:
\begin{equation}\label{eq7}
\epsilon_{m}k_{tz} + \epsilon_{t}k_{mz} \coth\left(\frac{\bar{k}_{mz}d}{2}\right) = 0\ ,
\end{equation}
\begin{equation}\label{eq8}
A_{t}^{a} = -2 A_0 \sinh\left(\frac{\bar{k}_{mz}d}{2}\right)\ .
\end{equation}
Now that we know the dispersion and amplitude relations for both the coupled modes, we can proceed to calculate explicit expressions for the electric and magnetic fields and thus calculate the spin. In SI units, we define the spin angular momentum density carried by the light, which gives the local expectation of the spin operator as \cite{berry2009,berry2015}
\begin{equation}\label{eq9}
\mathbf{s} = \frac{Im\left(\epsilon_{0}\mathbf{E^*}\times\mathbf{E} + \mu_{0}\mathbf{H^*}\times\mathbf{H}\right)}{4\omega}\ .
\end{equation}

\section{Numerical Calculations}
In this section we present the results of our numerical calculations. For our calculations we have used two wavelengths $\lambda=633$ nm and $\lambda=1550$ nm. We use $\epsilon_t =1$ and $\epsilon_m$ is calculated via spline interpolation from the data of Johnson and Christy \cite{johnson1972}.
But first, we have to deal with normalization in order to make our comparisons of the transverse spin for different wavelengths and metal film thicknesses $d$ meaningful. We normalize the power flow into the system at $x=0$ to unity
\begin{equation}\label{eq10}
\int_{-\infty}^{\infty}P_x dz = 1\ .
\end{equation}
Since spin is related to the spin part of the Poynting Vector as
\begin{equation}\label{eq11}
\mathbf{P}_s = c^2 \left(\nabla\times\mathbf{s}\right)\ ,
\end{equation}
the quantity
\begin{equation}\label{eq12}
\mathbf{s'}=\frac{\mathbf{s}c^2}{\int_{-\infty}^{\infty}P_x dz}\ ,
\end{equation}
is dimensionless. We integrate $P_x$ over $z$ as the coupled plasmons propagate in the $x$ direction localized in the metal film. Hereafter in this section,  we refer to this normalized, dimensionless $\mathbf{s'}$ as the transverse spin.
\par
We look first at the dispersion curves for the coupled surface plasmons for two different wavelengths: $\lambda = 1.55\ \mu$m and $\lambda = 0.633\ \mu$m.
\begin{figure}[t]
	\centering
	\includegraphics[width=1.0\linewidth]{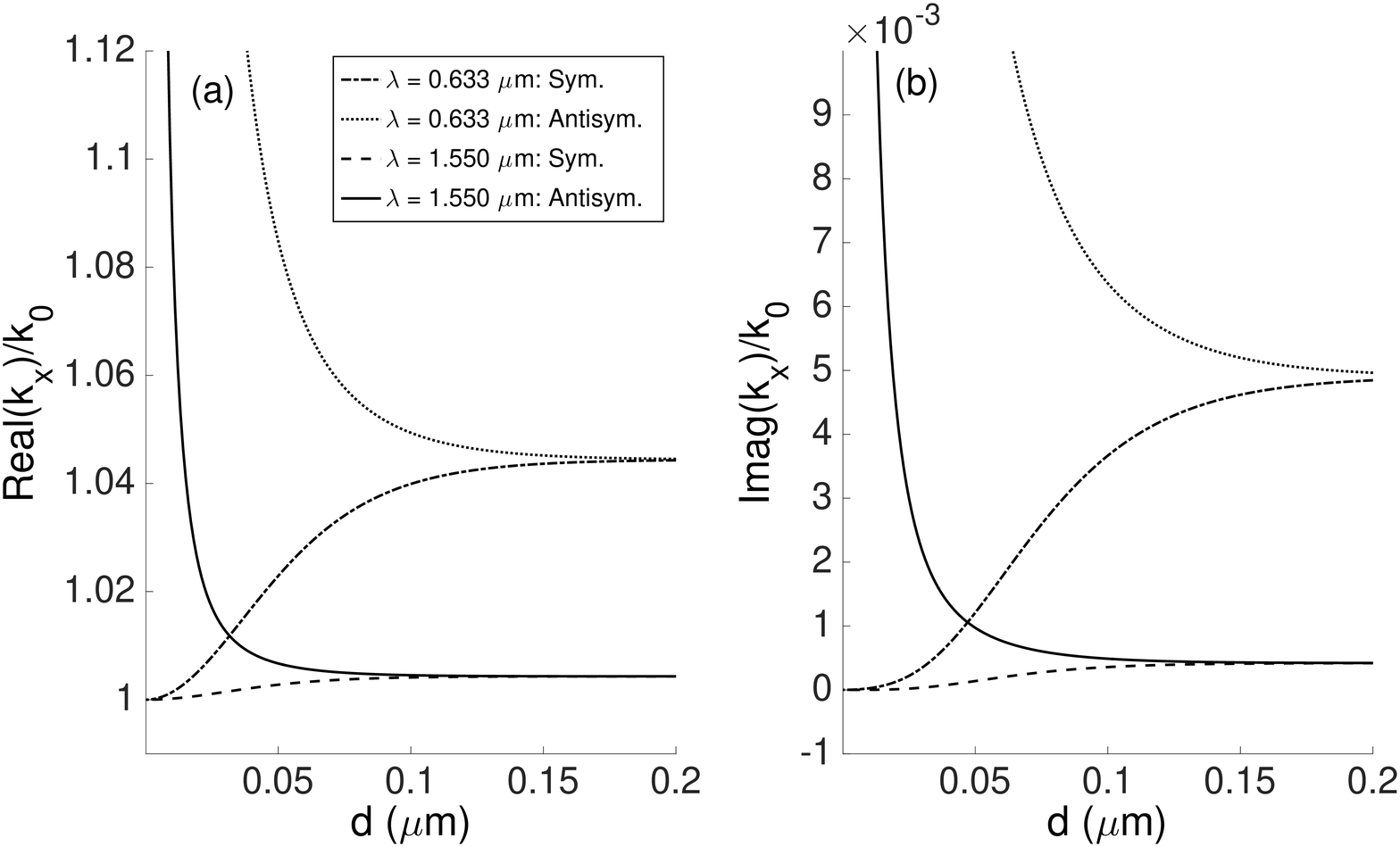}
	\caption{(a) \textbf{Dispersion curve}: Shows the real part of $k_x$ as a function of $d$. The solid line is for the antisymmetric mode $\lambda = 1.55\mu$m. The dashed line is for the symmetric mode at the same wavelength. The dotted line is for the antisymmetric mode at $\lambda = 0.633\mu$m. The dashed-dotted line is for the symmetric mode at the same wavelength. (b) is the same as (a) except that it shows the imaginary part of $k_x$.}
	\label{fig:Compare_Disp}
\end{figure}
figure \ref{fig:Compare_Disp} clearly shows that compared to $\lambda = 1.55\mu$m we get stronger coupling for a given $d$ at $\lambda = 0.633\mu$m. Also, from figure \ref{fig:Compare_Disp}(b) we can see that the antisymmetric mode has higher values for the imaginary part of $k_x$ compared to the symmetric mode. Thus the antisymmetric plasmon will decay faster than the symmetric plasmon while propagating along $x$. Thus the antisymmetric plasmon is called short range (SR) plasmon and the symmetric plasmon is called a long range (LR) plasmon.

As mentioned earlier, we are dealing with TM-polarized waves and the single magnetic field component has no contribution to the transverse spin. Any transverse spin would therefore have its origin in the rotation of the electric field. We now look at the electric field and how it behaves in the $x-z$ plane which is the plane of incidence. First we look at the fields for $\lambda = 0.633\ \mu$m.
\begin{figure}[h]
	\centering
	\includegraphics[width=0.8\linewidth]{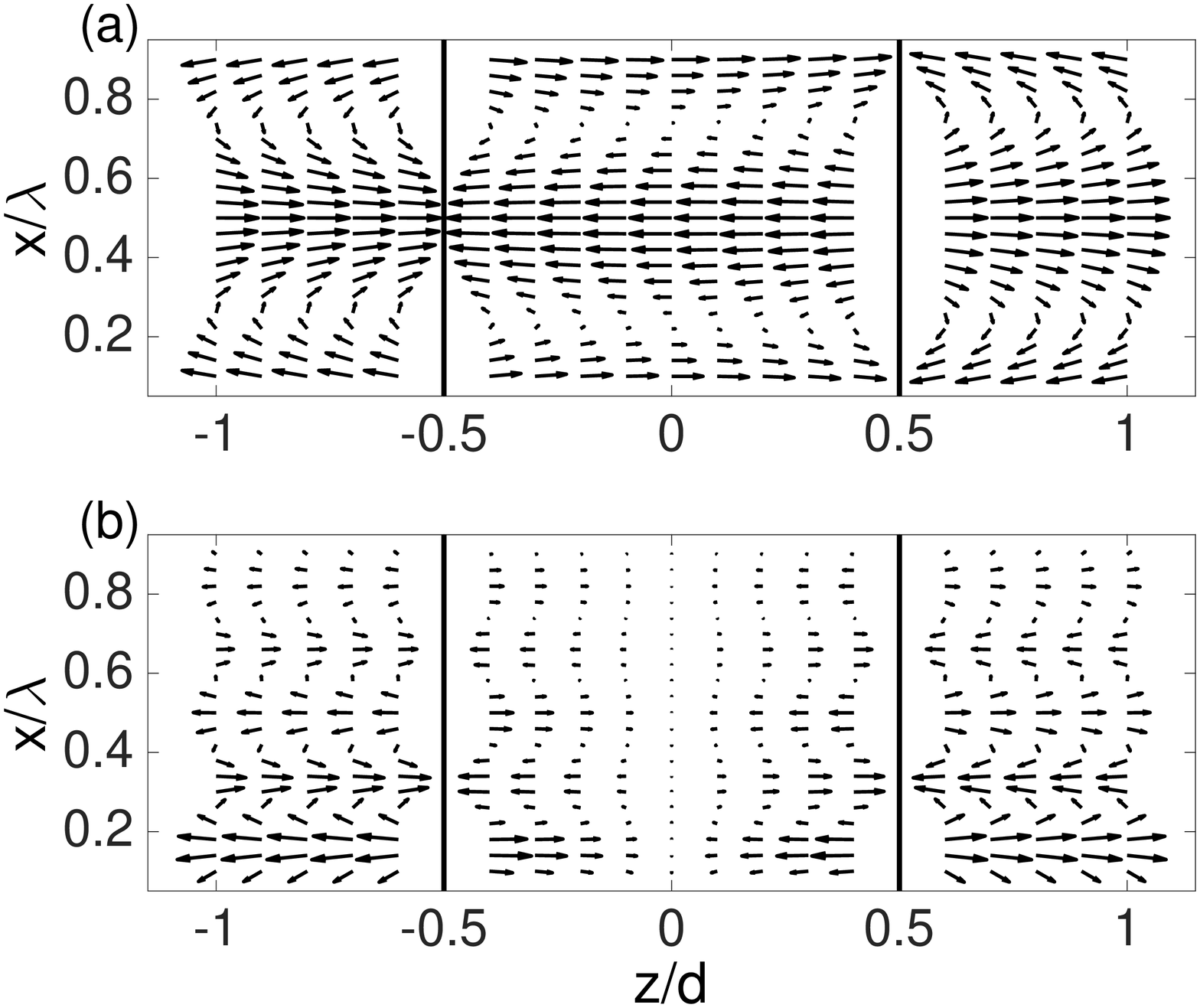}
	\caption{Electric Field in the plane of incidence for $\lambda = 0.633\ \mu$m and $d = 0.006\ \mu$m. (a) and (b) show the field distribution for the symmetric and antisymmetric modes, respectively.}
	\label{fig:Quiver_633}
\end{figure}
We see that as we move in the $x$ direction, the electric field rotates. The period of the rotation is much less for the antisymmetric mode compared to the symmetric mode. This is because of the higher value of the real part of $k_x$ in the antisymmetric mode. We also see that the field decays as we move along $x$. This is due to the high value of the imaginary part of $k_x$ in the antisymmetric mode.

\begin{figure}[t]
	\centering
	\includegraphics[width=0.8\linewidth]{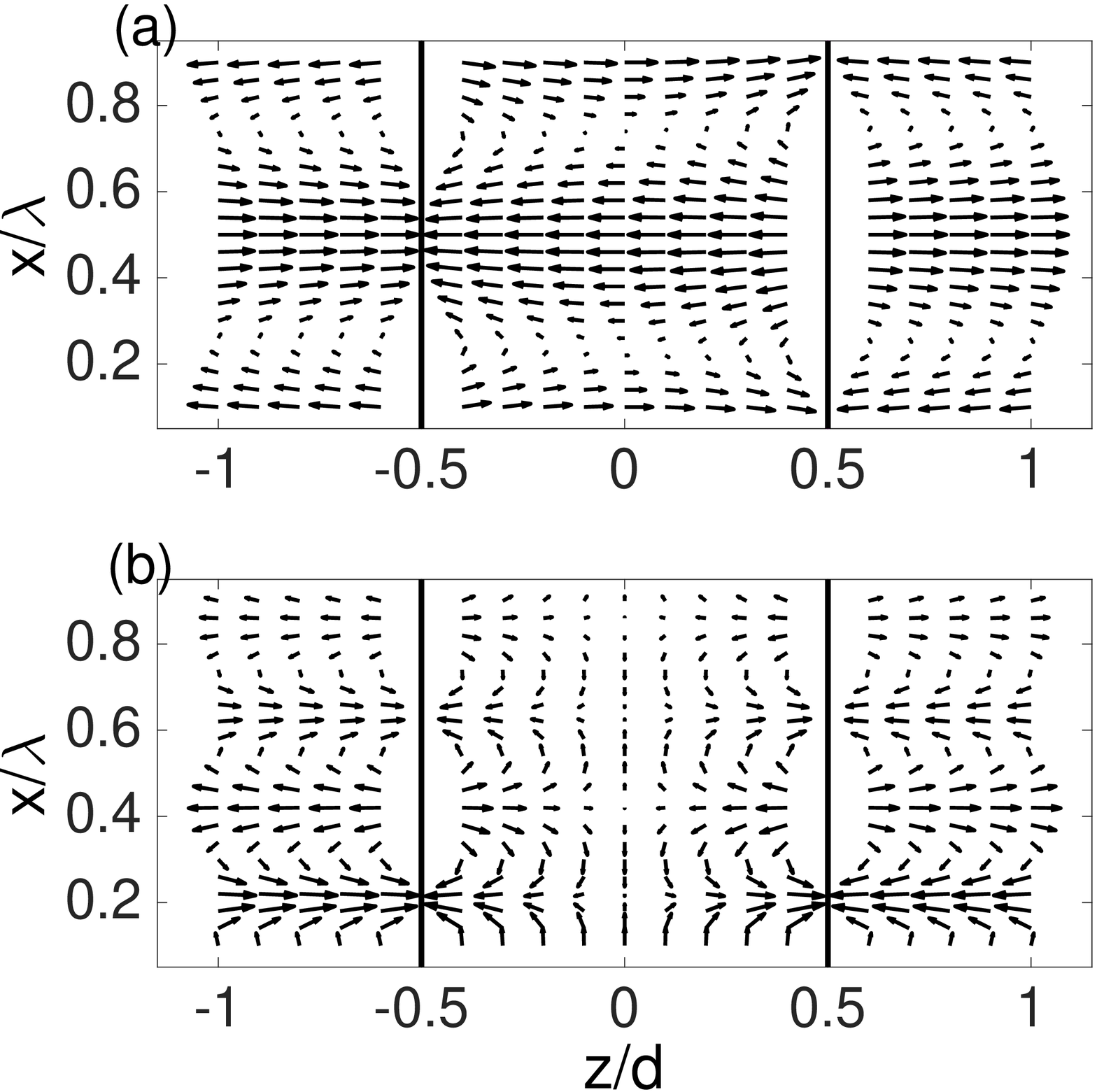}
	\caption{Electric Field in the plane of incidence for $\lambda = 1.55\ \mu$m and $d = 0.002\ \mu$m.  (a) and (b) show the field distribution for the symmetric and antisymmetric modes, respectively.}
	\label{fig:Quiver_1550}
\end{figure}

In figure \ref{fig:Quiver_1550} we look at the electric field for $\lambda = 1.55\ \mu$m and for $d=0.002\ \mu$m.
Note that we used a smaller thickness of metal ($d=0.002\ \mu$m) in order to have appreciable decay of the SR mode with $\lambda=1.55\ \mu$m as compared to $\lambda=0.633\ \mu$m (compare the scales of the imaginary part of $k_x$ in 
figures \ref{fig:Compare_Disp}(a) and \ref{fig:Compare_Disp}(b)). Again we see properties similar to figure \ref{fig:Quiver_633}. The field rotates faster in the antisymmetric case compared to the symmetric case and the fields also decay as we move along $x$. The reasons are again appreciable difference in the value of the real and imaginary parts of $k_x$ for the symmetric and antisymmetric modes.

Since we know that the rotation of the fields in a plane gives rise to the spin angular momentum along the axis of rotation, we realise that the rotation of the fields in the $x-z$ plane should give rise to a spin angular momentum along $y$. As shown in our analytic calculations for the lossless case in the next section, the numerical calculations for the lossy case also show that there is a transverse spin along the $y$ axis. Also, we must note here that the sense of rotation of the electric field changes across the metal-dielectric interface. Therefore the transverse spin density should change sign across the interface.

Now that we know to expect transverse spin from the electric field distribution, we look at the change in the transverse spin density of the modes as we change the strength of the coupling. The transverse spin is calculated at the metal-dielectrc interface.
\begin{figure}[t]
	\centering
	\includegraphics[width=1.0\linewidth]{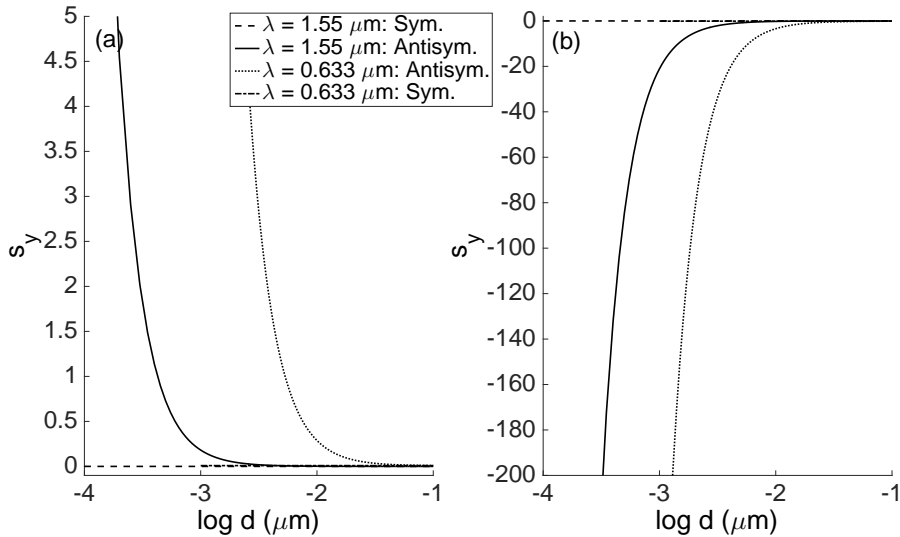}
	\caption{(a) Transverse spin density in the metal at $z=\frac{d}{2}$ as a function of metal film thickness $d$. The solid line is for the antisymmetric mode $\lambda = 1.55\mu$m. The dashed line is for the symmetric mode at the same wavelength. The dotted line is for the antisymmetric mode at $\lambda = 0.633\mu$m. The dashed-dotted line is for the symmetric mode at the same wavelength. (b) is the same as (a) but now the transverse spin in calculated in the dielectric.}
	\label{fig:Spin_comparison}
\end{figure}
We see that the transverse spin increases rapidly at small values of the metal film thickness $d$ for the antisymmetric (SR) mode. As expected, the transverse spin also changes sign across the metal-dielectric interface.

The behaviour of the transverse spin mimics thats of the dispersion curve. At small values of $d$, the transverse spin is distinct for the symmetric and antisymmetric modes. At larger values of $d$, in a manner akin to the dispersion curves, the spin for the symmetric and the antisymmetric modes become equal. Given that the transverse spin for the two cases is dependent on $k_x$, this behaviour may be expected and is clearly seen in figure \ref{fig:Spin_comparison}.
From figure \ref{fig:Spin_comparison} we note that the difference in the transverse spin density between the symmetric and antisymmetric modes increases with stronger coupling (smaller values of metal film thickness $d$). However, the difference is more pronounced at larger values of $d$ for smaller wavelengths. As stated earlier, from figure \ref{fig:Compare_Disp} we saw that the coupling is stronger for larger $d$ at smaller wavelength. Thus, we can conclude that the difference in the transverse spin between the coupled modes is directly related to the strength of the coupling. Stronger coupling leads to a greater difference in the transverse spin between the coupled modes.

\section{Analytical Results for the Lossless Case}
In this section we present explicit analytical results for having a better insight into the underlying physics. In order to be able to arrive at analytic expressions for the transverse spin angular momentum density, we use a specific assumption \cite{bliokh2012}. We assume that we are working with media, where the permittivitties under consideration are all real. As a consequence of ignoring ohmic losses in metals the calculations get drastically simplified with purely imaginary propagation constant in metal. It is to be noted that here we present results without any normalisation. Thus the results contain unknown constant amplitude $A_0$, which was earlier fixed by the normalization constraint given by eq. (\ref{eq11}).
\par
We start with the symmetric case. The electric and magnetic fields in the metal are as follows
\begin{equation}\label{eq13}
\mathbf{H}_m^s = 2 A_0 \cosh\left(\bar{k}_{mz}z\right)e^{ik_x x}\hat{y}\ ,
\end{equation}
\begin{equation}\label{eq14}
\mathbf{E}_m^s = \frac{A_0}{\omega\epsilon_{m}\epsilon_{0}} \left[-2i\bar{k}_{mz} \sinh\left(\bar{k}_{mz}z\right)\hat{x} - 2k_x \cosh\left(\bar{k}_{mz}z\right)\hat{z}\right] e^{ik_x x}\ .
\end{equation}
From eq. (\ref{eq13}) we can see that the magnetic field distribution inside the metal for the symmetric mode has $\cosh$ dependence on $z$ and thus is symmetric.\\
The electric and the magnetic fields in the dielectric can be expressed as
\begin{equation}\label{eq15}
\mathbf{H}_d^s = 2 A_0 \cosh\left(\frac{\bar{k}_{mz}d}{2}\right)e^{ik_x x} e^{-\bar{k}_{tz}\left(z-\frac{d}{2}\right)}\hat{y}\ ,
\end{equation}
\begin{equation} \label{eq16}
\mathbf{E}_m^s = \frac{2 A_0 \cosh\left(\frac{\bar{k}_{mz}d}{2}\right)}{\omega\epsilon_{t}\epsilon_{0}} \left[i\bar{k}_{tz}\hat{x} - k_x\hat{z}\right]e^{ik_x x}e^{-\bar{k}_{tz}\left(z-\frac{d}{2}\right)}\ .
\end{equation}
The point to be noted is the phase lag between the $x$ and $z$ component of the electric field vector (see eqs. (\ref{eq14}) and (\ref{eq16}) in the symmetric and antisymmetric modes. This creates a polarization ellipse in the plane of incidence and gives rise to a spin angular momentum perpendicular to it.

It is to be noted that since we are dealing with TM-polarized waves, there is no helicity dependent spin. The only spin we get is the aforementioned helicity independent transverse spin \cite{bliokh2012,bliokh2014,bliokh2015,bekshaev2015}. 
Because of the TM character of the incident fields there is no magnetic field contribution to the spin angular momentum density ($\mathbf{H}$ has only one non-zero component along the y-direction). Thus the only contribution to the transverse spin is due to the electric fields rotating in the plane of incidence due to the phase lag between the $x$ and $z$ components. following eq. (\ref{eq9}), one can write down the transverse spin density in the metal as
\begin{equation}\label{eq17}
\mathbf{s}_m^s = \frac{k_x \bar{k}_{mz}}{\omega^{3}\epsilon_{0}\epsilon_{m}^{2}} |A_0|^2\sinh\left(2\bar{k}_{mz}z\right)\hat{y}\ ,
\end{equation}
while transverse spin density in the dielectric is given by
\begin{equation}\label{eq18}
\mathbf{s}_d^s = \frac{-2|A_0|^2 \cosh^2\left(\frac{\bar{k}_{mz}d}{2}\right)}{\omega^{3}\epsilon_{0}\epsilon_{t}^{2}} k_x \bar{k}_{tz}e^{-2\bar{k}_{tz}\left(z-\frac{d}{2}\right)}\hat{y}\ .
\end{equation}
We now look at the antisymmetric case. The electric and the magnetic fields in the metal can be written as
\begin{equation} \label{eq19}
\mathbf{H}_m^a = -2 A_0\sinh\left(\bar{k}_{mz}z\right)e^{ik_x x}\hat{y}\ ,
\end{equation}
\begin{equation}\label{eq20}
\mathbf{E}_m^a = \frac{A_0}{\omega\epsilon_{m}\epsilon_{0}} \left[-2i\bar{k}_{mz} \cosh\left(\bar{k}_{mz}z\right)\hat{x} + 2k_x sinh\left(\bar{k}_{mz}z\right)\hat{z}\right]\\e^{ik_x x}\ .
\end{equation}
Similar expressions for the electric and the magnetic fields in dielectric are given by
\begin{equation}\label{eq21}
\mathbf{H}_d^a = -2 A_0 \sinh\left(\frac{\bar{k}_{mz}d}{2}\right)e^{ik_x x} e^{-\bar{k}_{tz}\left(z-\frac{d}{2}\right)}\hat{y}\ ,
\end{equation}
\begin{equation}\label{eq22}
\mathbf{E}_d^a = \frac{2 A_0 \sinh\left(\frac{\bar{k}_{mz}d}{2}\right)}{\omega\epsilon_{t}\epsilon_{0}} \left[-i\bar{k}_{tz}\hat{x} + k_x\hat{z}\right]e^{ik_x x}e^{-\bar{k}_{tz}\left(z-\frac{d}{2}\right)}\ .
\end{equation}
For the antisymmetric mode as well there is again a phase lag between the $x$ and $z$ components of the electric field (see eqs. (\ref{eq20}) and (\ref{eq22})) which leads to the transverse spin.
Finally the transverse spin density in the metal for the antisymmtric mode is given by
\begin{equation}\label{eq23}
\mathbf{s}_m^a = \frac{k_x \bar{k}_{mz}}{\omega^{3}\epsilon_{0}\epsilon_{m}^{2}} |A_0|^2\sinh\left(2\bar{k}_{mz}z\right)\hat{y}\ ,
\end{equation}
while for dielectric it can be written as
\begin{equation}\label{eq24}
\mathbf{s}_d^a = \frac{-2|A_0|^2\sinh^2\left(\frac{\bar{k}_{mz}d}{2}\right)}{\omega^{3}\epsilon_{0}\epsilon_{t}^{2}} k_x \bar{k}_{tz}e^{-2\bar{k}_{tz}\left(z-\frac{d}{2}\right)}\hat{y}\ .
\end{equation}
It is to be noted that even though the expressions for the transverse spin in the metal is the same for both the symmetric and the antisymmetric modes (see eqs. (\ref{eq17}) and (\ref{eq23})), it does not imply that the transverse spin is identical for the two cases. It depends in the value of $k_x$ for a given mode which is strongly dependent on the strength of coupling. Thus, for strong coupling (small values of $d$), there is a large difference in the value of $k_x$ (and consequently in $\mathbf{s_m}$ and $\mathbf{s_d}$) in the symmetric and antisymmetric modes (see figure \ref{fig:Compare_Disp}). The other important aspect that needs attention is the loss of distinction between the coupled modes in the context of propagation losses, since both of them are now lossless. In other words the nomenclature long- and short- range looses its meaning.

\section{Conclusion}
We have studied transverse spin in the context of coupled surface plasmons and show that the transverse spin is dependent on the symmetry and on the character of the modes, i.e., whether they are short range or long range. We note that for stronger coupling between the two interface plasmons, the difference in the transverse spin between the modes becomes more apparent. The transverse spin increases rapidly for small thicknesses in the short range mode for its more pronounced propagation losses. In contrast to some previous studies on transverse spin with single-interface plasmons, all numerical calculations are carried out retaining the loss effects in metal with the Johnson-Christie data. Analytical calculation for the lossless case clearly reveals the origin of the transverse spin in both the coupled modes. Our results agree with the fact that the transverse spin is dependent on the phase difference between the different components of the electric field. The short range mode has a large imaginary component of the wavevector and from the transversality condition, it follows that the imaginary part of the longitudinal component of the electric field will be large. Thus, we see large transverse spin for the short range mode. However, this large extraordinary transverse spin has to be measured at a very short distance within which the mode can survive.









\end{document}